\documentclass{article}
\pdfoutput=1 

\usepackage[font=small,skip=0pt]{caption}
\usepackage{algorithmic}
\usepackage{algorithm}
\usepackage{float}
\usepackage[caption = false]{subfig}
\makeatletter
\newcommand{\HEADER}[1]{\ALC@it\underline{\textsc{#1}}\begin{ALC@g}}
\newcommand{\ENDHEADER}{\end{ALC@g}}
\makeatother
 
\usepackage{microtype}
\usepackage{graphicx}
\usepackage{booktabs} 

\PassOptionsToPackage{hyphens}{url}\usepackage{hyperref}

\usepackage[accepted]{icml2024}

\usepackage{amsmath}
\usepackage{amssymb}
\usepackage{mathtools}
\usepackage{amsthm}

\usepackage[capitalize,noabbrev]{cleveref}

\theoremstyle{plain}
\newtheorem{theorem}{Theorem}[section]

\theoremstyle{definition}
\newtheorem{definition}[theorem]{Definition}

\theoremstyle{remark}

\icmltitlerunning{Validating Climate Models with SCWD}

\usepackage{natbib}
\renewcommand{\cite}[1]{\citet{#1}}

\begin{document}
\raggedbottom
\twocolumn[
\icmltitle{Validating Climate Models with Spherical Convolutional Wasserstein Distance}




\begin{icmlauthorlist}
\icmlauthor{Robert C. Garrett}{uiucs}
\icmlauthor{Trevor Harris}{tamu}
\icmlauthor{Bo Li}{uiucs}
\icmlauthor{Zhuo Wang}{uiuca}
\end{icmlauthorlist}

\icmlaffiliation{uiucs}{Department of Statistics, University of Illinois at Urbana-Champaign, Urbana, IL, USA}
\icmlaffiliation{uiuca}{Department of Atmospheric Sciences, University of Illinois at Urbana-Champaign, Urbana, IL, USA}
\icmlaffiliation{tamu}{Department of Statistics, Texas A\&M University, College Station, TX, USA}
\icmlcorrespondingauthor{Robert Garrett}{rcg4@illinois.edu}

\icmlkeywords{Climate Models, Wasserstein Distance, Convolution, Functional Data}

\vskip 0.3in
]



\printAffiliationsArxiv{} 

\begin{abstract}
The validation of global climate models is crucial to ensure the accuracy and efficacy of model output. We introduce the spherical convolutional Wasserstein distance to more comprehensively measure differences between climate models and reanalysis data. This new similarity measure accounts for spatial variability using convolutional projections and quantifies local differences in the distribution of climate variables. We apply this method to evaluate the historical model outputs of the Coupled Model Intercomparison Project (CMIP) members by comparing them to observational and reanalysis data products. Additionally, we investigate the progression from CMIP phase 5 to phase 6 and find modest improvements in the phase 6 models regarding their ability to produce realistic climatologies.
\end{abstract}

\section{Introduction}
\label{introduction}

\paragraph{Climate Model Validation} General Circulation Models, or climate models, are mathematical representations of the climate system that describe interactions between matter and energy through the ocean, atmosphere, and land \citep{washington2005introduction}. Climate models are the primary tool for investigating the response of the climate system to changes in forcing, such as increases in $\text{CO}_2$, and projecting future climate states \citep{flato2014evaluation}. 
To assess the plausibility of climate models, 
climate scientists compare output from model simulations against observational data \citep{rood2019validation}.
This comparison is the focus of climate model validation techniques for ensuring that climate models capture the dynamics of the climate system \citep{roca2021joint}.

The Coupled Model Intercomparison Project (CMIP) was initiated in 1995 as a comprehensive and systematic program for assessing climate models against each other and observational data \citep{cmip6experiments}. 
Each model in CMIP participates in a wide variety of experiments such as performing a historical simulation, a pre-industrial control simulation, and various simulations representing different scenarios for CO$_2$ emissions \citep{cmip6experiments}. 
Because historical simulations coincide with observational measurements, we can compare each model's synthetic climate distribution to the distribution of observational or quasi-observational data products \citep{raaisaanen2007reliable}, to assess their reconstructive skill.
For complete spatial coverage we compare against reanalysis data, a blend of observations and short-range weather forecasts through data assimilation \citep{bengtsson2004can}. This has become one popular climate model validation method \citep{flato2014evaluation}. 

\vspace{-2mm}
\paragraph{Previous approaches} Many statistical and machine learning-based methods have been applied to assess climate model output against reanalysis fields. 
The most common approach is to compute the $L2$ distance, or root mean square error (RMSE), between different features of the climate model output and the reanalysis field \citep{li2021comparative, zamani2020comparison, karim2020evaluation, ayugi2021comparison}. RMSE provides a direct measure of the differences between two climate fields but does not take into account randomness of data, so we should use it with caution in evaluating climate models.  
Another approach, which is invariant to bias, is to compute measures of correlation between climate model output and reanalysis fields \citep{zhao2021evaluation, zamani2020comparison, karim2020evaluation, ayugi2021comparison}.

More comprehensive approaches employ techniques for random processes to compare two spatial fields. For example, \cite{shen2002nonparametric} and \cite{cressie2008detecting} use the wavelet decomposition to compare the spatial-frequency content of two fields. \cite{hering2011comparing} measures the loss differential between models of spatial processes, \cite{lund2009revisiting} and \cite{li2012defining} compare the first and second moments of two random processes, and \cite{yun2022detection} identifies local differences in the mean and dependency structure between two spatiotemporal climate fields.  
Functional data analysis techniques have been introduced to compare spatial and spatiotemporal random fields, considering the random fields as continuous functions.
Many of these approaches compare the underlying mean functions from two sets of functional data \citep{zhang2007statistical, horvath2013estimation, staicu2014likelihood}. Other approaches include the second-order structure in the comparison \citep{zhang2015two, li2016comparison}, or compare the distributions of two spatial random processes \citep{harris2021evaluating}.

Since climate models aim to mimic the real climate which is the underlying pattern of weather, directly assessing the distributional differences between the modeled and observed data seems a more thorough approach to evaluating climate models.  \cite{vissio2020evaluating} proposed to use the Wasserstein distance (WD), a popular metric for comparing probability distributions \citep{villani2009wasserstein}, for such purpose. The WD can be intractable or even impossible to calculate between high-dimensional distributions \citep{kolouri2019generalized}, so \cite{vissio2020evaluating} first converts each climate field to a single spatial mean and then only compares the distribution of spatial means. However, this dimension reduction puts their method at the risk of missing important spatial variability information, and consequently failing to accurately distinguish two climate fields that are different.   

Recent contributions in the Machine Learning literature seek to compare multivariate distributions using many features while leveraging the efficiency of the one-dimensional WD \citep{vallender1974calculation}. The sliced WD \citep{bonneel2015sliced} compares random projections of distributions on $\mathbb{R}^n$, and the generalized sliced WD \citep{kolouri2019generalized} extends this to a broader class of projections. 
The convolutional sliced WD \citep{nguyen2022revisiting} compares distributions of discrete square images in the space $\mathbb{R}^{n\times n}$, accounting for the spatial structure using kernel convolutions. However, spatial fields from climate models are defined on a spherical domain, making them incompatible with the vectorized/square nature of these distances. The spherical sliced WD \citep{bonet2022spherical} can compare distributions of spatial point processes over a sphere, such as locations of natural disasters and extreme weather events, but cannot be used for smooth fields such as daily temperature and precipitation.
\vspace{-2mm}
\paragraph{Our proposal} We propose the functional sliced WD as a generalization of sliced WD to distributions of functional data. 
To create a tailored tool for climate model evaluation, we define the Spherical Convolutional Wasserstein distance (SCWD) as a special case of the functional sliced WD for functions on the unit sphere $\mathbb{S}^2$, the manifold on which latitude-longitude coordinates are defined. 
SCWD creates slices containing a small region of spatial fields to characterize local differences in the distribution of climate variables, which are further integrated into a single measure for global differences.  
Compared to the spatial mean-based WD from \cite{vissio2020evaluating}, SCWD accounts for spatial features while maintaining the modest computation for univariate WD when comparing climate models against reanalysis data, resulting in a comprehensive and more robust evaluation. 
We apply SCWD to rank climate models and assess the progression of the new CMIP era with respect to daily average surface temperature and daily total precipitation.

\section{Preliminaries}

We consider the problem of comparing two probability distributions $P$ and $Q$. 
Each distribution is a member of $\mathcal{P}(\Omega)$, the set of Borel probability measures on some sample space,  $\Omega$. 
In our application, we treat climate fields as functional data \citep{wang2016functional} over some spatial domain $\mathcal{S}$, thus we often assume $\Omega$ is a space of functions.
Our sample space of interest is $L^2(\mathcal{S})$, the set of square integrable functions from $\mathcal{S}\rightarrow\mathbb{R}$ where $\mathcal{S}$ is a compact subset of $\mathbb{R}^n$. 
Because CMIP model outputs are available at a global scale, we consider the spatial domain to be the unit sphere $\mathbb{S}^2$, the space over which latitude and longitude coordinates are assigned to the Earth's surface. 
In fact, the function space $L^2(\mathbb{S}^2)$ has been previously considered for modeling climate fields \citep{heaton2014constructing}.
To compare two functional data distributions $P$ and $Q$, we define a distance function $D(P,Q)$ to act as a similarity measure. 

\vspace{-2mm}
\paragraph{Comparing Distributions of Functions} 
Comparisons for distributions of functions in $L^2(\mathcal{S})$ have been studied for specific cases of $\mathcal{S}$ \citep{hall2007two, bugni2021permutation, pomann2016two, harris2021evaluating}, but none of these have focused on $L^2(\mathbb{S}^2)$.
The mathematical properties of  probability measures in $\mathcal{P}(L^2(\mathcal{S}))$ have been studied before \citep{gijbels2017general, kim2006invariant}, but it is challenging
to calculate distances such as the WD in this space without additional assumptions \citep{li2020functional}.
To define a similarity measure for distributions in $\mathcal{P}(L^2(\mathcal{S}))$, we build on the theory of the sliced WD and its various extensions, which have only been defined for distributions of finite-dimensional data.

\vspace{-2mm}
\paragraph{Sliced WD} 

Given a Borel function $\pi:\Omega\rightarrow\mathbb{R}$, the pushforward of $P$ under $\pi$ is a valid distribution in $\mathcal{P}(\mathbb{R})$ defined as $\pi\#P(B) = P(\pi^{-1}(B))$ for all Borel sets $B$ in $\mathbb{R}$. The $r$-th order sliced WD \citep{bonneel2015sliced} between $P,Q\in\mathcal{P}(\mathbb{R}^n)$ is a metric defined as the mean of the ordinary WD over univariate pushforwards:
\begin{equation}
\label{eq:slicedwasserstein}
    SW_r(P,Q) = \left(\int_{\mathbb{S}^{n-1}}W_r(\pi_\theta\#P,\pi_\theta\#Q)^rd\theta\right)^{1/r},
\end{equation}
where $\pi_\theta(x) = x^T\theta$ for $x\in\mathbb{R}^n$ and $\theta\in\mathbb{S}^{n-1}$. We call $\pi_\theta$ the slicing function because it produces one-dimensional ``slices'' of the data using projection matrices. Because $\pi_\theta\#P$ and $\pi_\theta\#Q$ are valid distributions in $\mathcal{P}(\mathbb{R})$, the WD inside the integral can be calculated with the commonly used analytical form for univariate measures \citep{vallender1974calculation}. Given two samples $X\sim P$ and $Y\sim Q$ we can approximate $SW_r(P,Q)$ using the following Monte Carlo integral:
\begin{equation}
\label{eq:mc}
    \left(\frac{1}{m}\sum_{i=1}^m\int_0^1\left|\hat{F}^{-1}_{X^T\theta_i}(q)-\hat{F}^{-1}_{Y^T\theta_i}(q)\right|^rdq\right)^{1/r},
\end{equation}
where $\theta_1,...,\theta_m$ is a random sample of projections from $\mathbb{S}^{n-1}$ and $\hat{F}^{-1}_{X^T\theta_i}$, $\hat{F}^{-1}_{Y^T\theta_i}$ are the empirical quantile functions of $X^T\theta_i$ and $Y^T\theta_i$, respectively, for $i=1,...,m$. 

\vspace{-2mm}
\paragraph{Generalized Sliced WD} The generalized sliced WD \citep{kolouri2019generalized} replaces $\pi_\theta$ with a more general class of slicing functions, denoted as $g_\theta$.
Because of this increased flexibility, the generalized sliced WD is a pseudometric rather than a metric except for specific cases of $g_\theta$ \citep{kolouri2019generalized}. 
For a given use case, the utility and metric properties of the generalized sliced WD are therefore determined by the choice of slicing function. Slicing functions can be chosen via optimization, estimated using neural networks, or specified by the researcher to isolate features of interest \citep{kolouri2019generalized}. The last option is desirable for our application, allowing the slices to be restricted to spatial features of interest to climate modelers.

However, the generalized sliced WD is defined between distributions of data in $\mathbb{R}^n$, a space that is not suitable for distributions of spatial fields. 
Climate fields could be coerced to vectors in $\mathbb{R}^n$ to be made compatible with the generalized sliced WD.
However, this would result in a loss of the inherent spatial structure, making it challenging to specify a slicing function that can handle considerations such as area weighting and spatial correlations.

\vspace{-2mm}
\paragraph{Convolution Sliced WD} The convolution sliced WD \citep{nguyen2022revisiting} represents images as matrices in the space $\mathbb{R}^{n\times n}$. The slicing function is replaced with kernel convolutions, or possibly a sequence of kernel convolutions, of $d\times d$ pixels for $d\in\mathbb{N}$. The kernel aggregates nearby locations to isolate local features, which could provide useful information to climate modelers. However, when climate fields are represented on a rectangular grid, the geographic area represented by each grid cell varies drastically between latitudes due to the non-Euclidean structure. Thus, a $d\times d$ pixel kernel will cover different sizes and shapes of geographic areas depending on location. For our application, the kernel radius should therefore be defined using geographic distance, not pixels.

\begin{figure*}[h]
\centering
\includegraphics[width=6.5in]{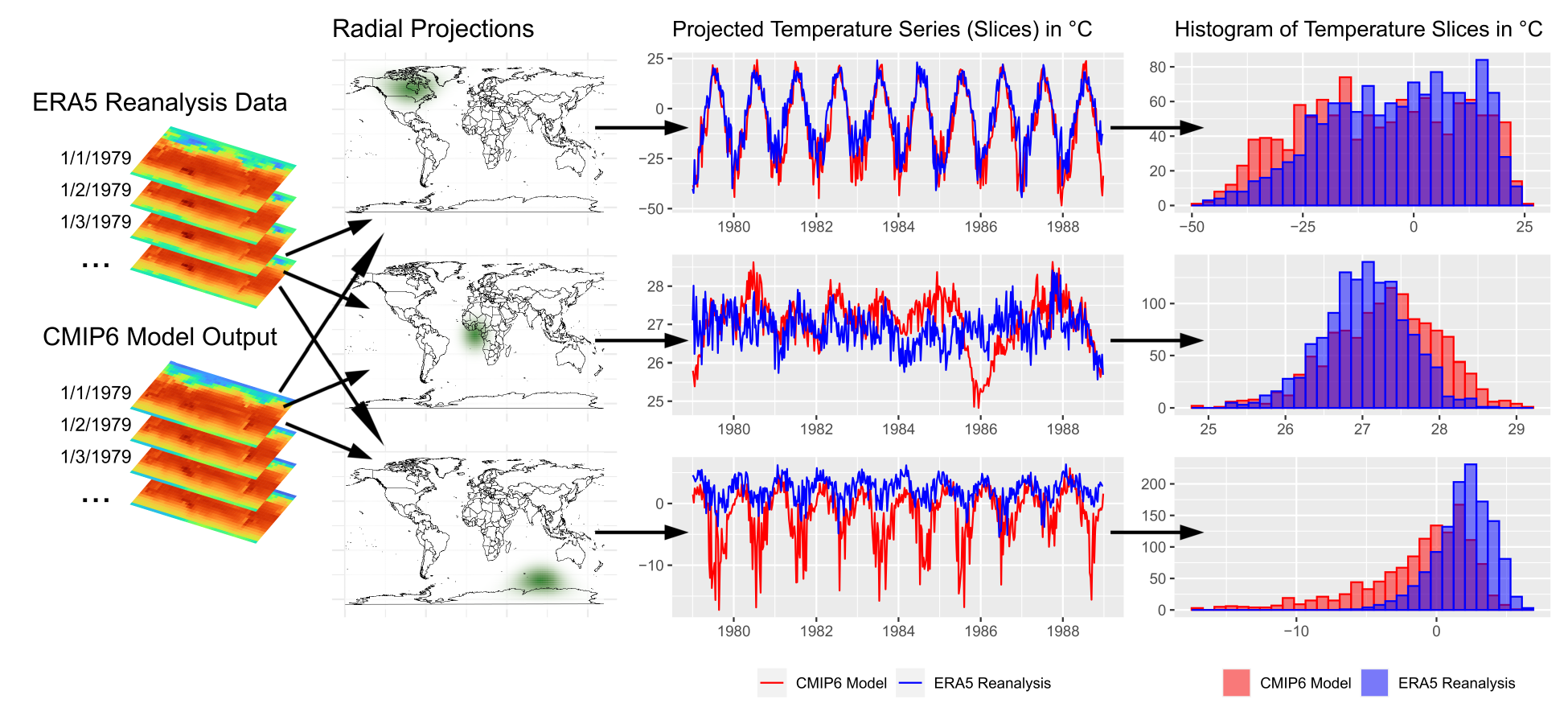}
\caption{Diagram representing the calculation of SCWD between two distributions of daily mean surface temperature fields from ERA5 and a CMIP6 model. For each day, radial projections are computed through kernel convolutions, represented here at three different locations. The resulting projections, called slices, summarise the local climate conditions in each dataset. The slices for each day are viewed as a sample from the marginal distribution at each location, represented here as histograms. SCWD is calculated as a global mean over the univariate WD between each pair of local distributions.}
\vspace{-2mm}
\label{fig:diagram}
\end{figure*}

\section{Methods}

We introduce the functional sliced WD framework which extends the flexible slicing process from the generalized sliced WD to the infinite-dimensional case of functions in $L^2(\mathcal{S})$. 
We focus on the special case of $L^2(\mathbb{S}^2)$, which we call the spherical convolutional WD (SCWD), for our climate model validation application. 
SCWD adapts the kernel convolution-based slicing idea of the convolutional sliced WD \citep{nguyen2022revisiting} to a continuous setting while accounting for the non-euclidean structure of climate fields realized over the Earth's surface. 


\subsection{Functional Sliced Wasserstein Distance}
\label{sec:fswd}


In general it is not possible to analytically characterize distributions in $\mathcal{P}(L^2(\mathcal{S}))$ and there are no closed form solutions for computing the Wasserstein distance. 
However, it is possible to slice elements of $L^2(\mathcal{S})$, meaning we can leverage the analytical form of the one-dimensional WD to define a computable sliced WD. 
We first extend the convolution slicer from \cite{nguyen2022revisiting} to the functional data case, allowing us to project functions in $L^2(\mathcal{S})$ to values in $\mathbb R$ while preserving local spatial information.
\begin{definition}[Convolution Slicer]
\label{def:slicer}
Let $\mathcal{S}$ be a compact subset of $\mathbb{R}^n$, $s\in\mathcal{S}$, and $k$, called the kernel function, be a continuous function from $\mathcal{S} \times \mathcal{S} \rightarrow [0,\infty)$.  We define the convolution slicer $c_s(f)$, a linear operator from $f\in L^2(\mathcal{S})\rightarrow\mathbb{R}$, as follows: 
\begin{equation*}
c_s(f) = \int_{\mathcal S} f(u)k(s,u)du.
\end{equation*}
\end{definition}

Since location $s\in\mathcal S$ is fixed, $k(s,u)$ is a continuous function from $u\in\mathcal{S}\rightarrow\mathbb{R}$. Because $\mathcal{S}$ is compact, it follows that $k(s,u)$ is bounded and thus $k(s,u) \in L^2(\mathcal{S})$. So, by Hölder's inequality, $c_s(f)$ is a bounded linear operator from $L^2(\mathcal{S})\rightarrow\mathbb{R}$. Bounded linear operators are also continuous \citep{stein2011functional}, so $c_s(f)$ is a continuous linear operator and, therefore, Borel measurable. Thus, for any measure $P\in\mathcal{P}(L^2(\mathcal{S}))$, the pushforward $c_s\#P$ is a valid measure in $\mathcal{P}(\mathbb{R})$. 
Thus, we can define a functional sliced WD between distributions in $\mathcal{P}(L^2(\mathcal{S}))$ as follows:

\begin{definition}[Functional Sliced WD]
\label{def:fswd}
Let $\mathcal{S}$ be a compact subset of $\mathbb{R}^n$, $r\geq 1$, and $P,Q\in\mathcal{P}(L^2(\mathcal{S}))$. Let $c_s$ be an operator satisfying Definition \ref{def:slicer}. We define the ($r$-th order) functional sliced WD between $P$ and $Q$ as follows:
\begin{equation*}
FSW_r(P,Q) = \left(\int_{\mathcal{S}}W_r(c_s\#P,c_s\#Q)^rds\right)^{1/r}
\end{equation*}
where $W_r$ is the  Wasserstein metric on $\mathcal{P}(\mathbb{R})$.
\end{definition}
Because $c_s\#P$ and $c_s\#Q$ are valid univariate probability measures, the analytical form of the univariate WD can be applied for efficient calculations. We introduce theoretical properties for the functional sliced WD in Theorem \ref{thm:pseudometric}

\begin{theorem}
\label{thm:pseudometric}
For all compact subsets $\mathcal{S}\subset\mathbb{R}^n$, $FSW_r$ is a pseudometric on $\mathcal{P}(\mathcal{F_S})$ and maintains the $r$-convexity property of the ordinary $W_r$ metric.
\end{theorem}
Proof of Theorem \ref{thm:pseudometric} is provided in Appendix \ref{app:pseudometric}. It is unknown if the final positivity property of a metric is satisfied. Proof of this property would require an invertible Radon-like transformation to be defined for probability measures in $\mathcal{P}(L^2(\mathcal{S}))$. 
Therefore, as with the generalized sliced WD, it is up to the researcher to specify an appropriate kernel function over the domain of interest. We provide such a choice for our application and give theoretical justification in Section \ref{sec:scwd}.

\subsection{Spherical Convolutional Wasserstein Distance}
\label{sec:scwd}

For our application to climate model validation, we specify the spatial domain $\mathcal S$ to be the unit sphere $\mathbb{S}^2$, the space over which latitude-longitude coordinates are assigned to locations on the Earth. To preserve local spatial information, we specify the slicing function to be a radial kernel function. We introduce the spherical convolutional WD (SCWD) as a specific case of the functional sliced WD: 
\begin{definition}[Spherical Convolutional WD]
\label{def:scwd}
Let $P,Q\in\mathcal{P}(L^2(\mathbb{S}^2))$ and $r\geq 1$. We define the ($r$-th order) SCWD between $P$ and $Q$ as follows:
\begin{equation*}
SCW_r(P,Q) = \left(\int_{\mathbb{S}^2}W_r(\omega_s\#P,\omega_s\#Q)^rds\right)^{1/r},
\end{equation*}
where $\omega_s$ is a convolution slicer that satisfies Definition \ref{def:slicer} with associated radial kernel function $\phi(s,u)$: 
\begin{equation*}
    \omega_s(f) = \int_{\mathbb{S}^2} f(u)\phi(s,u)du.
\end{equation*}
\end{definition}
Because $\phi$ is a radial kernel function, $\omega_s$ aggregates local information and the resulting pushforward measures $\omega_s\#P$ and $\omega_s\#Q$ represent the local distribution around each location $s$.
SCWD is therefore calculated as the global mean of the WD between local distributions at each location.
The local WD values can be recorded and later visualized as a map to pinpoint regions with higher or lower similarity.
Figure \ref{fig:diagram} demonstrates the process for calculating SCWD between two distributions of surface temperature fields, and details on the implementation are provided in Appendix \ref{app:implementation}. 
Similar to the sliced WD approximation (see equation \ref{eq:mc}), the computation uses Monte Carlo integration, with the outside sum replaced by a strided convolution over the sphere.

\vspace{-2mm}
\paragraph{Choice of Kernel} In our analysis, we specify the kernel function centered to be $\phi(|s-u|;l)$, where $|s-u|$ is the chordal distance between locations $s$ and $u$ and $\phi$ is the Wendland kernel function used in \cite{nychka2015multiresolution}:
\begin{equation}\label{eq:wendland}
    \phi(d;l) = 
    \begin{cases}
        (1-d)^6(35d^2+18d+3)/3 & d \leq l, \\
        0 & d > l,
    \end{cases}
\end{equation} 
with range parameter $l > 0$ determining the radius over which the kernel is nonzero. 
The Wendland kernel meets the continuity assumption in Definition \ref{def:slicer} and is also compact, which enables efficient sparse computations for our analysis.
 
Positive definite kernels, such as the Wendland kernel for $l$ less than the diameter \citep{hubbert2023generalised}, allow us to retain full spatial information via the spectral density.
This is because the convolution theorem on $\mathbb{S}^2$ \citep{driscoll1994computing} gives an injective correspondence between the spectral density of a function $f\in L^2(\mathbb{S}^2)$ and the spectral density of the convolution $(f*k)(s) = \int_{\mathbb{S}^2} f(u)k(s,u)du$ when $k$ is positive definite.
Note that as $l\rightarrow\infty$, the Wendland kernel converges to the flat kernel $\phi(d;\infty) = 1$, resulting in a SCWD where every slice is the global mean. 
In this case, the SCWD will be equal to the global mean-based WD from \cite{vissio2020evaluating}, leading to a complete loss of spatial variability information. 
In our analysis, we ensure a positive definite kernel by specifying $l$ to be less than the diameter of the Earth (about $12,750$ km).
We study the sensitivity of our results to this parameter in Section \ref{sec:comparison}.


\section{Climate Model Validation}
\label{sec:experiments}

We consider climate model outputs from the Coupled Model Intercomparison Project (CMIP) historical experiment phases 5 and 6. We focus on daily average near-surface (2m) temperature in degrees Celsius and daily total precipitation in mm.
The CMIP6 historical simulations are organized by ensembles, each of which is distinguished with an \texttt{ripf} identifier (\texttt{rip} for CMIP5), representing realization, initialization, physics, and forcings of the model, respectively  \citep{cmip6experiments}. We obtain 46 CMIP6 model outputs with the \texttt{r1i1p1f1} ID and 33 CMIP5 model outputs with the \texttt{r1i1p1} ID.
Output was obtained either at the daily frequency or aggregated from 3-hourly data. 
At the time of writing, two of the 79 total models did not have output available at a suitable frequency for surface temperature. 
All 79 models had output available for total precipitation.

To serve as references for climate model evaluation, we collect the European Centre for Medium-Range Weather Forecasts (ECMWF) Reanalysis 5th Generation (ERA5) \citep{hersbach2020era5} as well as the Reanalysis-2 data from the National Centers for Environmental Protection (NCEP) \citep{kanamitsu2002ncep} reanalysis datasets. Both datasets were obtained for surface temperature and total precipitation at a daily frequency. Due to known issues with reanalysis data for precipitation \citep{tapiador2017global}, we obtain observations from the National Centers for Environmental Information (NCEI) Global Precipitation Climatology Project (GPCP) Daily Precipitation Analysis Climate Data Record \citep{huffman2001global,adler2020global} as an additional reference for precipitation only. 

The historical time periods for each climate variable were chosen to maximize the available model outputs and reference datasets.
For surface temperature, a common time period of January 1, 1979 to November 30th, 2005 was collected for each CMIP output and reanalysis dataset. 
For total precipitation, we restrict the time period to October 1, 1996 to November 30th, 2005 to accommodate the first available day of observations in the GPCP dataset. 
Each data product represents the climate variables on a different latitude-longitude grid, which varies in size and structure. 
See Appendix \ref{sec:data} for full details on the spatial resolution and availability of temperature/precipitation for each dataset.

\subsection{Evaluation of CMIP6 Models}

\begin{figure*}[h]
\centering
\includegraphics[width=5.6in]{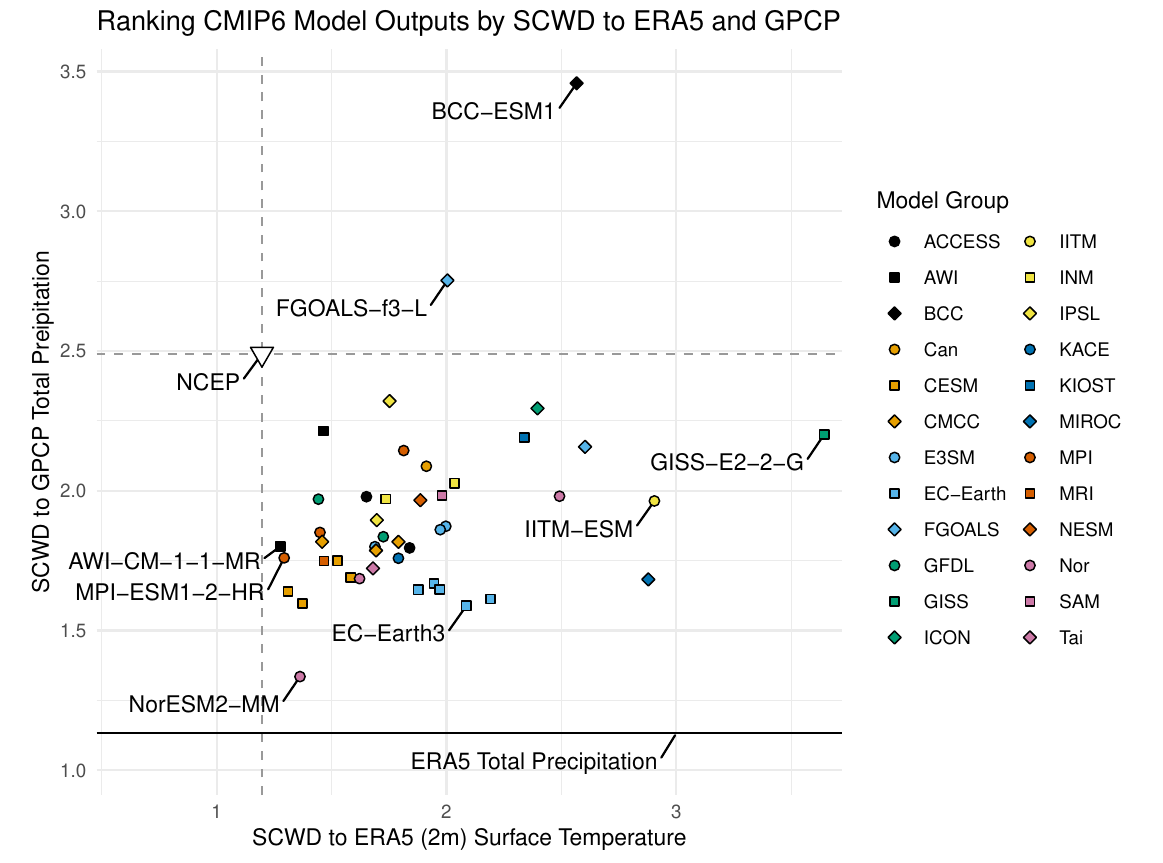}
\caption{Ranking CMIP6 model outputs using SCWD. Each model output is represented by a point on the scatter plot and models from the same group share the color and shape. The x-axis and y-axis values represent each model's SCWD to the ERA5 surface temperature and GPCP total precipitation fields, respectively. The NCEP reanalysis is included as a blank triangle with dashed lines representing the SCWD to ERA5 and GPCP. The SCWD from the ERA5 total precipitation field to GPCP is represented as a solid line.}
\vspace{-2mm}
\label{fig:cmip6_rankings}
\end{figure*}

To evaluate the skill of CMIP6 models in characterizing local climate distributions, we compute the SCWD between each model output and the observed/reanalysis datasets. 
Smaller SCWD values indicate more similarity between local distributions in a model output and the reference dataset.
We do not expect perfect agreement between models and historical data, so, similar to \cite{vissio2020evaluating}, we additionally calculate SCWD between the reference datasets as a baseline for comparison. All SCWD calculations in this section are performed using the Wendland kernel with range parameter of 1,000km for slicing. 

We select the ERA5 Reanalysis as our reference dataset for (2m) surface temperature due to its high spatial resolution. We calculate distances from each surface temperature model output in our CMIP5 and CMIP6 ensembles to ERA5. In addition, we calculate the SCWD between ERA5 and the NCEP Reanalysis to compare the variability between reanalysis datasets to the variability between models and reanalysis. For total precipitation, we select the GPCP observational data as our reference dataset. We calculate distances from each total precipitation model output in our CMIP5 and CMIP6 ensembles to GPCP. 
We include SCWD calculations from GPCP to both ERA5 and NCEP for comparison, with the secondary goal of assessing the accuracy of each reanalysis in faithfully filling gaps in observed precipitation measurements. Full details on the SCWD rankings can be seen in Tables \ref{tab:cmip6} and \ref{tab:cmip5} in Appendix \ref{app:tables}. Here, we focus on the results for CMIP6, and Figure \ref{fig:cmip6_rankings} provides the SCWD  from each CMIP6 model output to the ERA5 surface temperature field and GPCP total precipitation field. 

\vspace{-2mm}
\paragraph{Surface Temperature} For surface temperature, NCEP has a lower SCWD to ERA5 than all CMIP models. This is not a surprise: both ERA5 and NCEP are based on observations, so we expect their temperature distributions to be similar in most regions. Among the model outputs, many models have a SCWD to ERA5 similar to that of NCEP. In particular, AWI-CM-1-1-MR from the Alfred Wegener Institute and MPI-ESM1-2-HR from the Max Planck Institute have the lowest SCWD for surface temperature, with a few models close behind. 

\vspace{-2mm}
\paragraph{Total Precipitation} Compared to surface temperature, NCEP no longer has a lower SCWD to GPCP than the CMIP6 models.
Instead, the ERA5 total precipitation field, which has the lowest SCWD to GPCP, serves as a better baseline for comparison.
Deficiencies of precipitation from reanalysis have been reported in previous studies \citep{janowiak1998comparison}. In brevity, precipitation is sensitive to model physics and is not strongly constrained by observations via data assimilation. The low SCWD of ERA5 can likely be attributed to the high model resolution and more advanced model physics and data assimilation system compared to NCEP.
Among the CMIP6 models, the Norwegian Climate Centre NorESM2-MM model output stands out with the lowest SCWD to GPCP by a relatively wide margin. The EC-Earth3 and CESM model outputs also form clusters with low SCWD values for precipitation. 

No single model has the lowest SCWD value for both surface temperature and total precipitation, but NorESM2-MM seems to have the best balance of low distances for each variable and is not far off from the intersection of the lines for our reference datasets. Similarly, no model has the highest SCWD value for both climate variables. The GISS-E2-2-G model from NASA's Goddard Institute for Space Studies is a distinct outlier with a high surface temperature SCWD to ERA5, and the Beijing Climate Center BCC-ESM1 model is an outlier in terms of high SCWD to GPCP. We investigate these high SCWD values in Section \ref{sec:spatial}.

\subsection{Spatial Comparisons}
\label{sec:spatial}

\begin{figure*}[h]
\centering
\subfloat{\includegraphics[width = 6.5in]{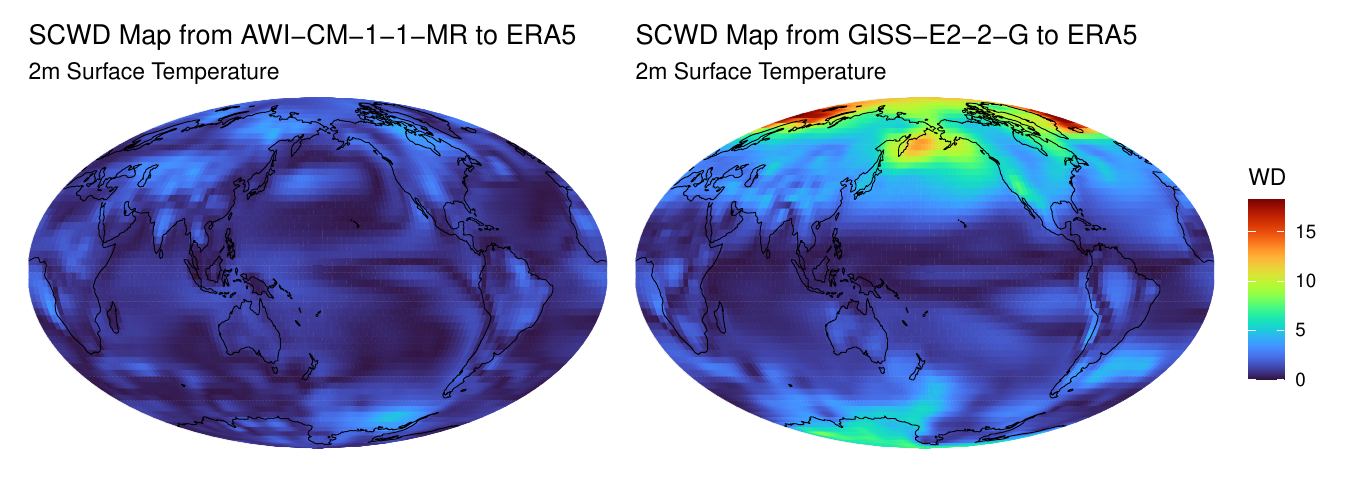}} \\
\subfloat{\includegraphics[width = 6.5in]{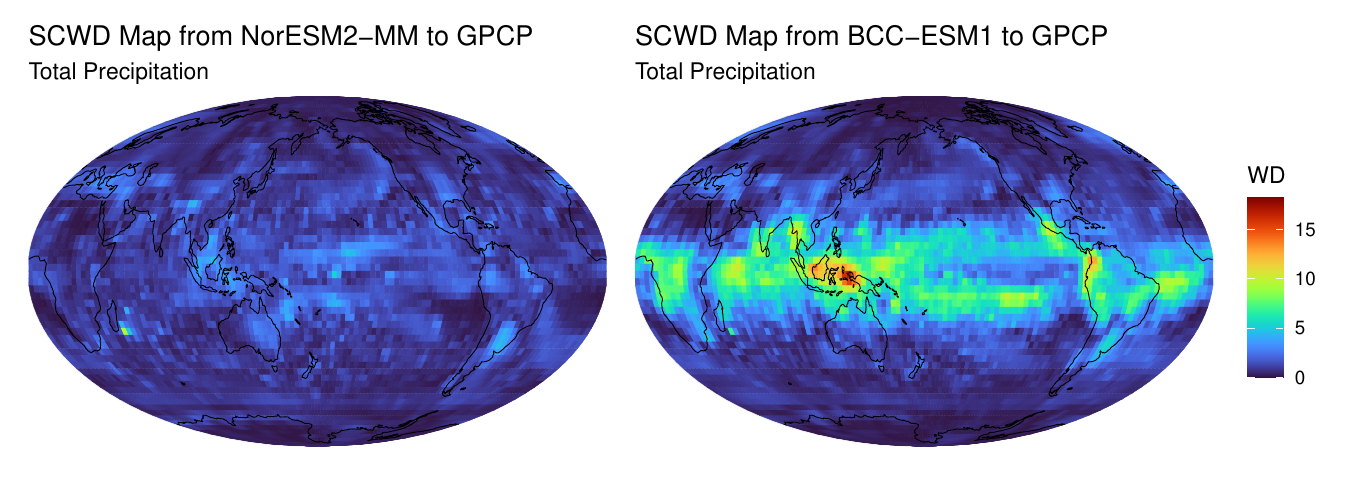}}
\vspace{-2mm}
\caption{Top: Map of local Wasserstein distances from ERA5 to two CMIP6 2m surface temperature outputs: AWI-CM-1-1-MR and GISS-E2-2-G. 
Bottom: Map of local Wasserstein distances from GPCP to two CMIP6 total precipitation outputs: NorESM2-MM and BCC-ESM1.
Color fill at each location is determined by the WD between the local distributions obtained from the convolution slicer in Definition \ref{def:scwd}. The color scale is shared for both maps and continental boundaries are included in black to aid spatial comparisons.}
\label{fig:maps}
\end{figure*}


Because SCWD is calculated as a global mean of local WD values, we can investigate the geographic sources of these outlying high SCWD values. Figure \ref{fig:maps} provides a spatial breakdown of the local WD values obtained when calculating SCWD for surface temperature between ERA5 and AWI-CM-1-1-MR as well as ERA5 and GISS-E2-2-G, the models with the lowest and highest SCWD to ERA5, respectively. Overall, both maps seem smooth or continuous in space, with little variation between neighboring locations in most cases. For the map between AWI-CM-1-1-MR and ERA5, the local WD values are relatively low everywhere, with regions of slightly higher values near the poles and mountains. Compared to AWI-CM-1-1-MR, the GISS-E2-2-G model has similarly low WD values in the tropics. However, closer to the poles, the local WD values begin to increase. In particular, the Arctic region has extremely high WD values relative to the rest of the Earth. This is indicative of the previously documented winter cool bias in the Artic region for GISS-E2-2-G \citep{kelley2020giss}.


Similar maps are provided for the NorESM2-MM and BCC-ESM1 outputs compared to GPCP total precipitation. 
Overall, both maps are less smooth than the surface temperature maps, potentially due to the more localized nature of precipitation. 
Looking at both models, the higher values of SCWD near the equator in the Pacific and Atlantic oceans may be related to the double-Intertropical Convergence Zone problem common in CMIP models, in which excessive precipitation is produced in the southern tropics \citep{mechoso1995seasonal}. This trend is much more pronounced for BCC-ESM1 than for NorESM2-MM. Additionally, BCC-ESM1 has a region of particularly high WD values around eastern Indonesia, where wind-terrain interaction plays an important role in the regional distribution of precipitation. This region has been previously highlighted in \cite{zhang2021bcc} as an area where BCC-ESM1 heavily overestimates annual mean precipitation.

\subsection{Comparing CMIP5 and CMIP6}

\begin{figure}[h]
\centering
\includegraphics[width=2.8in]{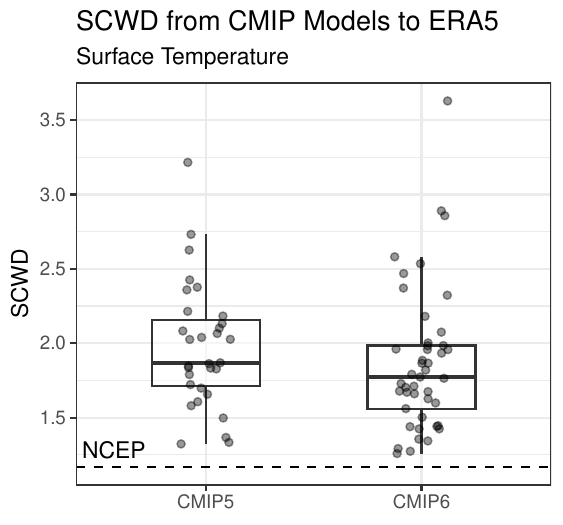}
\vspace{-1mm}
\caption{Boxplots of SCWD from the CMIP5 and CMIP6 model outputs to the ERA5 Reanalysis for 2m surface temperature. Each point represents the SCWD value for one climate model output to ERA5, with CMIP5 and CMIP6 separated into two boxplots for comparison. A dashed line is included to represent the SCWD from the NCEP Reanalysis to ERA5.}
\vspace{-2mm}
\label{fig:scwd_tas}
\end{figure}

Figure \ref{fig:scwd_tas} provides the SCWD calculations for each CMIP model and NCEP to the ERA5 Reanalysis.
Between the two CMIP eras, the median SCWD value for CMIP6 models to ERA5 is lower than that of CMIP5. However, the two boxplots are overall very similar, so the difference between the CMIP5 and CMIP6 ensembles is subtle. In addition, the model with the highest SCWD to ERA5 comes from the CMIP6 ensemble. Overall, we see a promising, albeit limited, decrease in SCWD for general CMIP6 models compared to CMIP5. This indicates improved performance of CMIP6 when it comes to reconstructing realistic temperature distributions at the local level.

\begin{figure}[h]
\centering
\includegraphics[width=2.8in]{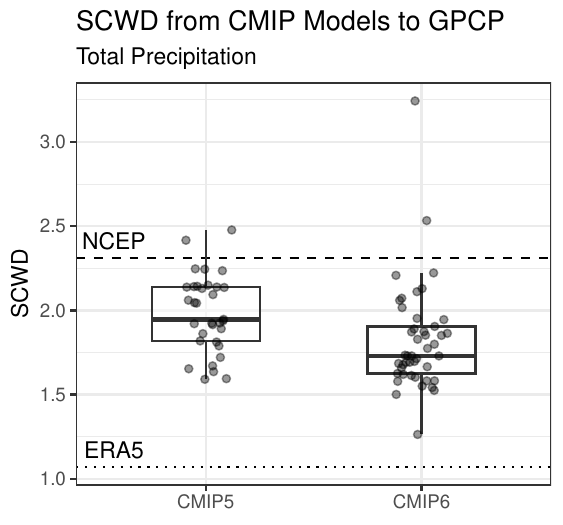}
\vspace{-1mm}
\caption{Boxplots of SCWD from the CMIP5 and CMIP6 model outputs to the GPCP observational dataset for total precipitation. Each point represents the SCWD value for one climate model output to GPCP, with CMIP5 and CMIP6 separated into two boxplots for comparison. Dotted and dashed lines are included to represent the SCWD from ERA5 and NCEP to GPCP, respectively.}
\label{fig:scwd_pr}
\vspace{-2mm}
\end{figure}

Figure \ref{fig:scwd_pr} provides the SCWD calculations for each CMIP model and ERA5/NCEP to the GPCP observations.
Between the two CMIP phases, the median SCWD value for CMIP6 is again lower than that of CMIP5. Compared to the surface temperature boxplots in Figure \ref{fig:scwd_tas}, the difference between CMIP5 and CMIP6 is even more distinct. This indicates that relative to surface temperature, precipitation modeling has seen greater gains with the transition from CMIP5 to CMIP6. 
The improvement of the CMIP6 models compared to CMIP5 in precipitation representation has been reported in some previous studies using different evaluation methods (e.g., \cite{chen2021evaluation}). It can be probably attributed to the more advanced model physics and overall higher model resolution in CMIP6.

\subsection{Metric Comparison}
\label{sec:comparison}

We compare our method to a previous climate model validation approach and assess our choice of range parameter by generating three additional rankings for the CMIP model outputs: SCWD with a range of 500km and 2,500km respectively, and the global mean-based WD from \cite{vissio2020evaluating}. These rankings can be seen alongside our original 1,000km SCWD results in Appendix \ref{app:tables}. Here, we focus on the CMIP6 results in Table \ref{tab:cmip6} in the Appendix.

Overall, the SCWD rankings are nearly identical regardless of the range parameter. The only exception is for a few total precipitation models, such as GFDL-CM4 and E3SM-2-0, which see improvement in the rankings for higher range parameters. However, there is a larger discrepancy between the rankings for SCWD with 1,000km range and the global mean-based WD (GMWD). For surface temperature, even though the SCWD from NCEP to ERA5 is lower than that of all CMIP6 models, the GMWD ranks NCEP among the median CMIP6 models. For total precipitation, we see a similar result for ERA5 compared to GPCP: ERA5 has the best ranking when using SCWD but only an average ranking for GMWD. 
These results indicate that incorporating local perspectives, rather than just the global mean, leads to a distance that is more skilled at determining similarities between climate models and observational datasets. More specifically, a model that represents the global mean well may have large compensating errors at the regional scale. SCWD can better represent the regional performance of climate models than GMWD and the individual slices provide more information about where a model excels or struggles.





\section{Discussion and Future Work}

Climate model validation is critical for ensuring that climate models faithfully represent the Earth system. To this end, we developed a new similarity measure, called spherical convolutional Wasserstein distance (SCWD), which quantifies model performance in a way that properly accounts for spatial variability. SCWD builds on previous sliced WD methods to compare distributions of infinite-dimensional functional data, specifically surfaces on the sphere $\mathbb{S}^2$. We show improved performance over previous approaches for model evaluation: SCWD better recognizes the similarities between observed and reanalysis datasets relative to climate models. In addition, SCWD identifies regions where climate models deviate from observed conditions, providing useful insights to climate modelers.

Due to the shared spherical domain, SCWD could be directly applied to compare distributions of $\text{360}^o$ images.
In this case, a straightforward extension would allow for multiple convolution layers similar to the rectangular case in \cite{nguyen2022revisiting}.  
SCWD could also be adapted to compare distributions of spatiotemporal fields, rather than spatial fields as in this article, by including both time and space in the functional data domain.
This would require specifying a space-time kernel function \citep{porcu2016spatio}.
The more general functional sliced WD framework has a variety of potential use cases for comparing distributions of functions on a broad class of manifolds. 
Potential application areas where the data lie on non-trivial manifolds include facial recognition \citep{li2014differential}, astronomy \citep{szapudi2008introduction}, and ecology \citep{sutherland2015modelling}.


\clearpage

\section{Acknowledgements}

This research was partially supported by the National Science Foundation through awards DGE-1922758 and DMS-2124576.

\section{Impact Statement}

Climate models are the primary tool for climate prediction and projection. However, deficiencies persist among climate models despite decades of improvement. Our method offers a succinct yet informative approach to evaluating climate models against observational and reanalysis data, both on the global and regional scales. It provides valuable guidance for climate model improvement and can assist in selecting more reliable models to reduce uncertainty in climate projections, which is crucial for climate mitigation and adaptation efforts.


\bibliography{ICML}

\begin{thebibliography}{57}
\providecommand{\natexlab}[1]{#1}
\providecommand{\url}[1]{\texttt{#1}}
\expandafter\ifx\csname urlstyle\endcsname\relax
  \providecommand{\doi}[1]{doi: #1}\else
  \providecommand{\doi}{doi: \begingroup \urlstyle{rm}\Url}\fi

\bibitem[Adler et~al.(2020)Adler, Wang, Sapiano, Huffman, Bolvin, Nelkin,
  et~al.]{adler2020global}
Adler, R., Wang, J.-J., Sapiano, M., Huffman, G., Bolvin, D., Nelkin, E.,
  et~al.
\newblock Global precipitation climatology project (gpcp) climate data record
  (cdr), version 1.3 (daily).
\newblock 2020.

\bibitem[Ayugi et~al.(2021)Ayugi, Zhihong, Zhu, Ngoma, Babaousmail, Rizwan, and
  Dike]{ayugi2021comparison}
Ayugi, B., Zhihong, J., Zhu, H., Ngoma, H., Babaousmail, H., Rizwan, K., and
  Dike, V.
\newblock Comparison of cmip6 and cmip5 models in simulating mean and extreme
  precipitation over east africa.
\newblock \emph{International Journal of Climatology}, 41\penalty0
  (15):\penalty0 6474--6496, 2021.

\bibitem[Bengtsson et~al.(2004)Bengtsson, Hagemann, and
  Hodges]{bengtsson2004can}
Bengtsson, L., Hagemann, S., and Hodges, K.~I.
\newblock Can climate trends be calculated from reanalysis data?
\newblock \emph{Journal of Geophysical Research: Atmospheres}, 109\penalty0
  (D11), 2004.

\bibitem[Bonet et~al.(2022)Bonet, Berg, Courty, Septier, Drumetz, and
  Pham]{bonet2022spherical}
Bonet, C., Berg, P., Courty, N., Septier, F., Drumetz, L., and Pham, M.-T.
\newblock Spherical sliced-wasserstein.
\newblock \emph{arXiv preprint arXiv:2206.08780}, 2022.

\bibitem[Bonneel et~al.(2015)Bonneel, Rabin, Peyr{\'e}, and
  Pfister]{bonneel2015sliced}
Bonneel, N., Rabin, J., Peyr{\'e}, G., and Pfister, H.
\newblock Sliced and radon wasserstein barycenters of measures.
\newblock \emph{Journal of Mathematical Imaging and Vision}, 51:\penalty0
  22--45, 2015.

\bibitem[Bugni \& Horowitz(2021)Bugni and Horowitz]{bugni2021permutation}
Bugni, F.~A. and Horowitz, J.~L.
\newblock Permutation tests for equality of distributions of functional data.
\newblock \emph{Journal of Applied Econometrics}, 36\penalty0 (7):\penalty0
  861--877, 2021.

\bibitem[Chen et~al.(2021)Chen, Hsu, and Liang]{chen2021evaluation}
Chen, C.-A., Hsu, H.-H., and Liang, H.-C.
\newblock Evaluation and comparison of cmip6 and cmip5 model performance in
  simulating the seasonal extreme precipitation in the western north pacific
  and east asia.
\newblock \emph{Weather and Climate Extremes}, 31:\penalty0 100303, 2021.

\bibitem[Cressie et~al.(2008)Cressie, Pavlicov{\'a}, and
  Santner]{cressie2008detecting}
Cressie, N., Pavlicov{\'a}, M., and Santner, T.~J.
\newblock Detecting signals in fmri data using powerful fdr procedures.
\newblock \emph{Statistics and its interface}, 1\penalty0 (1):\penalty0 23--32,
  2008.

\bibitem[Driscoll \& Healy(1994)Driscoll and Healy]{driscoll1994computing}
Driscoll, J.~R. and Healy, D.~M.
\newblock Computing fourier transforms and convolutions on the 2-sphere.
\newblock \emph{Advances in applied mathematics}, 15\penalty0 (2):\penalty0
  202--250, 1994.

\bibitem[Eyring et~al.(2016)Eyring, Bony, Meehl, Senior, Stevens, Stouffer, and
  Taylor]{cmip6experiments}
Eyring, V., Bony, S., Meehl, G.~A., Senior, C.~A., Stevens, B., Stouffer,
  R.~J., and Taylor, K.~E.
\newblock Overview of the coupled model intercomparison project phase 6 (cmip6)
  experimental design and organization.
\newblock \emph{Geoscientific Model Development}, 9\penalty0 (5):\penalty0
  1937--1958, 2016.
\newblock \doi{10.5194/gmd-9-1937-2016}.
\newblock URL \url{https://gmd.copernicus.org/articles/9/1937/2016/}.

\bibitem[Flato et~al.(2014)Flato, Marotzke, Abiodun, Braconnot, Chou, Collins,
  Cox, Driouech, Emori, Eyring, et~al.]{flato2014evaluation}
Flato, G., Marotzke, J., Abiodun, B., Braconnot, P., Chou, S.~C., Collins, W.,
  Cox, P., Driouech, F., Emori, S., Eyring, V., et~al.
\newblock Evaluation of climate models.
\newblock In \emph{Climate change 2013: the physical science basis.
  Contribution of Working Group I to the Fifth Assessment Report of the
  Intergovernmental Panel on Climate Change}, pp.\  741--866. Cambridge
  University Press, 2014.

\bibitem[Gijbels \& Nagy(2017)Gijbels and Nagy]{gijbels2017general}
Gijbels, I. and Nagy, S.
\newblock On a general definition of depth for functional data.
\newblock 2017.

\bibitem[Hall \& Van~Keilegom(2007)Hall and Van~Keilegom]{hall2007two}
Hall, P. and Van~Keilegom, I.
\newblock Two-sample tests in functional data analysis starting from discrete
  data.
\newblock \emph{Statistica Sinica}, pp.\  1511--1531, 2007.

\bibitem[Harris et~al.(2021)Harris, Li, Steiger, Smerdon, Narisetty, and
  Tucker]{harris2021evaluating}
Harris, T., Li, B., Steiger, N.~J., Smerdon, J.~E., Narisetty, N., and Tucker,
  J.~D.
\newblock Evaluating proxy influence in assimilated paleoclimate
  reconstructions—testing the exchangeability of two ensembles of spatial
  processes.
\newblock \emph{Journal of the American Statistical Association}, 116\penalty0
  (535):\penalty0 1100--1113, 2021.

\bibitem[Heaton et~al.(2014)Heaton, Katzfuss, Berrett, and
  Nychka]{heaton2014constructing}
Heaton, M., Katzfuss, M., Berrett, C., and Nychka, D.
\newblock Constructing valid spatial processes on the sphere using kernel
  convolutions.
\newblock \emph{Environmetrics}, 25\penalty0 (1):\penalty0 2--15, 2014.

\bibitem[Hering \& Genton(2011)Hering and Genton]{hering2011comparing}
Hering, A.~S. and Genton, M.~G.
\newblock Comparing spatial predictions.
\newblock \emph{Technometrics}, 53\penalty0 (4):\penalty0 414--425, 2011.

\bibitem[Hersbach et~al.(2020)Hersbach, Bell, Berrisford, Hirahara,
  Hor{\'a}nyi, Mu{\~n}oz-Sabater, Nicolas, Peubey, Radu, Schepers,
  et~al.]{hersbach2020era5}
Hersbach, H., Bell, B., Berrisford, P., Hirahara, S., Hor{\'a}nyi, A.,
  Mu{\~n}oz-Sabater, J., Nicolas, J., Peubey, C., Radu, R., Schepers, D.,
  et~al.
\newblock The era5 global reanalysis.
\newblock \emph{Quarterly Journal of the Royal Meteorological Society},
  146\penalty0 (730):\penalty0 1999--2049, 2020.

\bibitem[Horv{\'a}th et~al.(2013)Horv{\'a}th, Kokoszka, and
  Reeder]{horvath2013estimation}
Horv{\'a}th, L., Kokoszka, P., and Reeder, R.
\newblock Estimation of the mean of functional time series and a two-sample
  problem.
\newblock \emph{Journal of the Royal Statistical Society: Series B (Statistical
  Methodology)}, 75\penalty0 (1):\penalty0 103--122, 2013.

\bibitem[Hubbert \& J{\"a}ger(2023)Hubbert and
  J{\"a}ger]{hubbert2023generalised}
Hubbert, S. and J{\"a}ger, J.
\newblock Generalised wendland functions for the sphere.
\newblock \emph{Advances in Computational Mathematics}, 49\penalty0
  (1):\penalty0 3, 2023.

\bibitem[Huffman et~al.(2001)Huffman, Adler, Morrissey, Bolvin, Curtis, Joyce,
  McGavock, and Susskind]{huffman2001global}
Huffman, G.~J., Adler, R.~F., Morrissey, M.~M., Bolvin, D.~T., Curtis, S.,
  Joyce, R., McGavock, B., and Susskind, J.
\newblock Global precipitation at one-degree daily resolution from
  multisatellite observations.
\newblock \emph{Journal of hydrometeorology}, 2\penalty0 (1):\penalty0 36--50,
  2001.

\bibitem[Janowiak et~al.(1998)Janowiak, Gruber, Kondragunta, Livezey, and
  Huffman]{janowiak1998comparison}
Janowiak, J.~E., Gruber, A., Kondragunta, C., Livezey, R.~E., and Huffman,
  G.~J.
\newblock A comparison of the ncep--ncar reanalysis precipitation and the gpcp
  rain gauge--satellite combined dataset with observational error
  considerations.
\newblock \emph{Journal of Climate}, 11\penalty0 (11):\penalty0 2960--2979,
  1998.

\bibitem[Kanamitsu et~al.(2002)Kanamitsu, Ebisuzaki, Woollen, Yang, Hnilo,
  Fiorino, and Potter]{kanamitsu2002ncep}
Kanamitsu, M., Ebisuzaki, W., Woollen, J., Yang, S.-K., Hnilo, J., Fiorino, M.,
  and Potter, G.
\newblock Ncep--doe amip-ii reanalysis (r-2).
\newblock \emph{Bulletin of the American Meteorological Society}, 83\penalty0
  (11):\penalty0 1631--1644, 2002.

\bibitem[Karim et~al.(2020)Karim, Tan, Ayugi, Babaousmail, and
  Liu]{karim2020evaluation}
Karim, R., Tan, G., Ayugi, B., Babaousmail, H., and Liu, F.
\newblock Evaluation of historical cmip6 model simulations of seasonal mean
  temperature over pakistan during 1970--2014.
\newblock \emph{Atmosphere}, 11\penalty0 (9):\penalty0 1005, 2020.

\bibitem[Kelley et~al.(2020)Kelley, Schmidt, Nazarenko, Bauer, Ruedy, Russell,
  Ackerman, Aleinov, Bauer, Bleck, et~al.]{kelley2020giss}
Kelley, M., Schmidt, G.~A., Nazarenko, L.~S., Bauer, S.~E., Ruedy, R., Russell,
  G.~L., Ackerman, A.~S., Aleinov, I., Bauer, M., Bleck, R., et~al.
\newblock Giss-e2. 1: Configurations and climatology.
\newblock \emph{Journal of Advances in Modeling Earth Systems}, 12\penalty0
  (8):\penalty0 e2019MS002025, 2020.

\bibitem[Kim(2006)]{kim2006invariant}
Kim, J.~U.
\newblock Invariant measures for a stochastic nonlinear schr{\"o}dinger
  equation.
\newblock \emph{Indiana University mathematics journal}, pp.\  687--717, 2006.

\bibitem[Kolouri et~al.(2019)Kolouri, Nadjahi, Simsekli, Badeau, and
  Rohde]{kolouri2019generalized}
Kolouri, S., Nadjahi, K., Simsekli, U., Badeau, R., and Rohde, G.
\newblock Generalized sliced wasserstein distances.
\newblock \emph{Advances in neural information processing systems}, 32, 2019.

\bibitem[Li \& Smerdon(2012)Li and Smerdon]{li2012defining}
Li, B. and Smerdon, J.~E.
\newblock Defining spatial comparison metrics for evaluation of paleoclimatic
  field reconstructions of the common era.
\newblock \emph{Environmetrics}, 23\penalty0 (5):\penalty0 394--406, 2012.

\bibitem[Li et~al.(2016)Li, Zhang, and Smerdon]{li2016comparison}
Li, B., Zhang, X., and Smerdon, J.~E.
\newblock Comparison between spatio-temporal random processes and application
  to climate model data.
\newblock \emph{Environmetrics}, 27\penalty0 (5):\penalty0 267--279, 2016.

\bibitem[Li et~al.(2021)Li, Huo, Chen, Zhao, and Zhao]{li2021comparative}
Li, J., Huo, R., Chen, H., Zhao, Y., and Zhao, T.
\newblock Comparative assessment and future prediction using cmip6 and cmip5
  for annual precipitation and extreme precipitation simulation.
\newblock \emph{Frontiers in Earth Science}, 9, 2021.
\newblock ISSN 2296-6463.
\newblock \doi{10.3389/feart.2021.687976}.
\newblock URL
  \url{https://www.frontiersin.org/article/10.3389/feart.2021.687976}.

\bibitem[Li et~al.(2014)Li, Turaga, Srivastava, and
  Chellappa]{li2014differential}
Li, R., Turaga, P., Srivastava, A., and Chellappa, R.
\newblock Differential geometric representations and algorithms for some
  pattern recognition and computer vision problems.
\newblock \emph{Pattern Recognition Letters}, 43:\penalty0 3--16, 2014.

\bibitem[Li \& Ma(2020)Li and Ma]{li2020functional}
Li, T. and Ma, J.
\newblock Functional data clustering analysis via the learning of gaussian
  processes with wasserstein distance.
\newblock In \emph{Neural Information Processing: 27th International
  Conference, ICONIP 2020, Bangkok, Thailand, November 23--27, 2020,
  Proceedings, Part II 27}, pp.\  393--403. Springer, 2020.

\bibitem[Lund \& Li(2009)Lund and Li]{lund2009revisiting}
Lund, R. and Li, B.
\newblock Revisiting climate region definitions via clustering.
\newblock \emph{Journal of Climate}, 22\penalty0 (7):\penalty0 1787--1800,
  2009.

\bibitem[Mechoso et~al.(1995)Mechoso, Robertson, Barth, Davey, Delecluse, Gent,
  Ineson, Kirtman, Latif, Le~Treut, et~al.]{mechoso1995seasonal}
Mechoso, C.~R., Robertson, A.~W., Barth, N., Davey, M., Delecluse, P., Gent,
  P., Ineson, S., Kirtman, B., Latif, M., Le~Treut, H., et~al.
\newblock The seasonal cycle over the tropical pacific in coupled
  ocean--atmosphere general circulation models.
\newblock \emph{Monthly Weather Review}, 123\penalty0 (9):\penalty0 2825--2838,
  1995.

\bibitem[Nguyen \& Ho(2022)Nguyen and Ho]{nguyen2022revisiting}
Nguyen, K. and Ho, N.
\newblock Revisiting sliced wasserstein on images: From vectorization to
  convolution.
\newblock \emph{Advances in Neural Information Processing Systems},
  35:\penalty0 17788--17801, 2022.

\bibitem[Nychka et~al.(2015)Nychka, Bandyopadhyay, Hammerling, Lindgren, and
  Sain]{nychka2015multiresolution}
Nychka, D., Bandyopadhyay, S., Hammerling, D., Lindgren, F., and Sain, S.
\newblock A multiresolution gaussian process model for the analysis of large
  spatial datasets.
\newblock \emph{Journal of Computational and Graphical Statistics}, 24\penalty0
  (2):\penalty0 579--599, 2015.

\bibitem[Pomann et~al.(2016)Pomann, Staicu, and Ghosh]{pomann2016two}
Pomann, G.-M., Staicu, A.-M., and Ghosh, S.
\newblock A two sample distribution-free test for functional data with
  application to a diffusion tensor imaging study of multiple sclerosis.
\newblock \emph{Journal of the Royal Statistical Society. Series C, Applied
  Statistics}, 65\penalty0 (3):\penalty0 395, 2016.

\bibitem[Porcu et~al.(2016)Porcu, Bevilacqua, and Genton]{porcu2016spatio}
Porcu, E., Bevilacqua, M., and Genton, M.~G.
\newblock Spatio-temporal covariance and cross-covariance functions of the
  great circle distance on a sphere.
\newblock \emph{Journal of the American Statistical Association}, 111\penalty0
  (514):\penalty0 888--898, 2016.

\bibitem[Ra{\"a}isa{\"a}nen(2007)]{raaisaanen2007reliable}
Ra{\"a}isa{\"a}nen, J.
\newblock How reliable are climate models?
\newblock \emph{Tellus A: Dynamic Meteorology and Oceanography}, 59\penalty0
  (1):\penalty0 2--29, 2007.

\bibitem[Roca et~al.(2021)Roca, Haddad, Akimoto, Alexander, Behrangi, Huffman,
  Kato, Kidd, Kirstetter, Kubota, et~al.]{roca2021joint}
Roca, R., Haddad, Z.~S., Akimoto, F.~F., Alexander, L., Behrangi, A., Huffman,
  G., Kato, S., Kidd, C., Kirstetter, P.-E., Kubota, T., et~al.
\newblock The joint ipwg/gewex precipitation assessment, 2021.

\bibitem[Rood(2019)]{rood2019validation}
Rood, R.~B.
\newblock \emph{Validation of Climate Models: An Essential Practice}, pp.\
  737--762.
\newblock Springer International Publishing, Cham, 2019.
\newblock ISBN 978-3-319-70766-2.
\newblock \doi{10.1007/978-3-319-70766-2_30}.
\newblock URL \url{https://doi.org/10.1007/978-3-319-70766-2_30}.

\bibitem[Shen et~al.(2002)Shen, Huang, and Cressie]{shen2002nonparametric}
Shen, X., Huang, H.-C., and Cressie, N.
\newblock Nonparametric hypothesis testing for a spatial signal.
\newblock \emph{Journal of the American Statistical Association}, 97\penalty0
  (460):\penalty0 1122--1140, 2002.

\bibitem[Staicu et~al.(2014)Staicu, Li, Crainiceanu, and
  Ruppert]{staicu2014likelihood}
Staicu, A.-M., Li, Y., Crainiceanu, C.~M., and Ruppert, D.
\newblock Likelihood ratio tests for dependent data with applications to
  longitudinal and functional data analysis.
\newblock \emph{Scandinavian Journal of Statistics}, 41\penalty0 (4):\penalty0
  932--949, 2014.

\bibitem[Stein \& Shakarchi(2011)Stein and Shakarchi]{stein2011functional}
Stein, E.~M. and Shakarchi, R.
\newblock \emph{Functional analysis: introduction to further topics in
  analysis}, volume~4.
\newblock Princeton University Press, 2011.

\bibitem[Sutherland et~al.(2015)Sutherland, Fuller, and
  Royle]{sutherland2015modelling}
Sutherland, C., Fuller, A.~K., and Royle, J.~A.
\newblock Modelling non-euclidean movement and landscape connectivity in highly
  structured ecological networks.
\newblock \emph{Methods in Ecology and Evolution}, 6\penalty0 (2):\penalty0
  169--177, 2015.

\bibitem[Szapudi(2008)]{szapudi2008introduction}
Szapudi, I.
\newblock Introduction to higher order spatial statistics in cosmology.
\newblock In \emph{Data Analysis in Cosmology}, pp.\  457--492. Springer, 2008.

\bibitem[Tapiador et~al.(2017)Tapiador, Navarro, Levizzani, Garc{\'\i}a-Ortega,
  Huffman, Kidd, Kucera, Kummerow, Masunaga, Petersen,
  et~al.]{tapiador2017global}
Tapiador, F., Navarro, A., Levizzani, V., Garc{\'\i}a-Ortega, E., Huffman, G.,
  Kidd, C., Kucera, P., Kummerow, C., Masunaga, H., Petersen, W., et~al.
\newblock Global precipitation measurements for validating climate models.
\newblock \emph{Atmospheric Research}, 197:\penalty0 1--20, 2017.

\bibitem[Vallender(1974)]{vallender1974calculation}
Vallender, S.
\newblock Calculation of the wasserstein distance between probability
  distributions on the line.
\newblock \emph{Theory of Probability \& Its Applications}, 18\penalty0
  (4):\penalty0 784--786, 1974.

\bibitem[Villani \& Villani(2009)Villani and Villani]{villani2009wasserstein}
Villani, C. and Villani, C.
\newblock The wasserstein distances.
\newblock \emph{Optimal Transport: Old and New}, pp.\  93--111, 2009.

\bibitem[Vissio et~al.(2020)Vissio, Lembo, Lucarini, and
  Ghil]{vissio2020evaluating}
Vissio, G., Lembo, V., Lucarini, V., and Ghil, M.
\newblock Evaluating the performance of climate models based on wasserstein
  distance.
\newblock \emph{Geophysical Research Letters}, 47\penalty0 (21):\penalty0
  e2020GL089385, 2020.

\bibitem[Wang et~al.(2016)Wang, Chiou, and M{\"u}ller]{wang2016functional}
Wang, J.-L., Chiou, J.-M., and M{\"u}ller, H.-G.
\newblock Functional data analysis.
\newblock \emph{Annual Review of Statistics and its application}, 3:\penalty0
  257--295, 2016.

\bibitem[Washington \& Parkinson(2005)Washington and
  Parkinson]{washington2005introduction}
Washington, W.~M. and Parkinson, C.
\newblock \emph{Introduction to three-dimensional climate modeling}.
\newblock University science books, 2005.

\bibitem[Yun et~al.(2022)Yun, Zhang, and Li]{yun2022detection}
Yun, S., Zhang, X., and Li, B.
\newblock Detection of local differences in spatial characteristics between two
  spatiotemporal random fields.
\newblock \emph{Journal of the American Statistical Association}, 117\penalty0
  (537):\penalty0 291--306, 2022.

\bibitem[Zamani et~al.(2020)Zamani, Hashemi~Monfared, Hamidianpour,
  et~al.]{zamani2020comparison}
Zamani, Y., Hashemi~Monfared, S.~A., Hamidianpour, M., et~al.
\newblock A comparison of cmip6 and cmip5 projections for precipitation to
  observational data: the case of northeastern iran.
\newblock \emph{Theoretical and Applied Climatology}, 142\penalty0
  (3):\penalty0 1613--1623, 2020.

\bibitem[Zhang et~al.(2021)Zhang, Wu, Zhang, Furtado, Xin, Shi, Li, Chu, Zhang,
  Liu, et~al.]{zhang2021bcc}
Zhang, J., Wu, T., Zhang, F., Furtado, K., Xin, X., Shi, X., Li, J., Chu, M.,
  Zhang, L., Liu, Q., et~al.
\newblock Bcc-esm1 model datasets for the cmip6 aerosol chemistry model
  intercomparison project (aerchemmip).
\newblock \emph{Advances in Atmospheric Sciences}, 38:\penalty0 317--328, 2021.

\bibitem[Zhang \& Chen(2007)Zhang and Chen]{zhang2007statistical}
Zhang, J.-T. and Chen, J.
\newblock Statistical inferences for functional data.
\newblock 2007.

\bibitem[Zhang \& Shao(2015)Zhang and Shao]{zhang2015two}
Zhang, X. and Shao, X.
\newblock Two sample inference for the second-order property of temporally
  dependent functional data.
\newblock 2015.

\bibitem[Zhao et~al.(2021)Zhao, He, Dong, Zhou, Xie, Mei, Wan, and
  Jiang]{zhao2021evaluation}
Zhao, S., He, W., Dong, T., Zhou, J., Xie, X., Mei, Y., Wan, S., and Jiang, Y.
\newblock Evaluation of the performance of cmip5 models to simulate land
  surface air temperature based on long-range correlation.
\newblock \emph{Frontiers in Environmental Science}, pp.\ ~6, 2021.

\end{thebibliography}
\bibliographystyle{icml2024}

\newpage
\appendix
\onecolumn
\section{Proof of Theorem \ref{thm:pseudometric}}
\label{app:pseudometric}
\begin{proof}
Let $r\geq 1$, let $\mathcal{S}$ be a compact subset of $\mathbb{R}^n$, and let $P,Q,U \in \mathcal{P}(L^2(\mathcal{S}))$. We show the identity property holds:
\begin{align*}
FSW_r(P,P) &= \left(\int_{\mathcal{S}}W_r(c_s\#P,c_s\#P)^rds\right)^{1/r}\\
&= \left(\int_{\mathcal{S}}0^rds\right)^{1/r}\\ 
&= 0
\end{align*}
The second line holds by the identity property of the ordinary WD. Next, the symmetry property:
\begin{align*}
FSW_r(P,Q) &= \left(\int_{\mathcal{S}}W_r(c_s\#P,c_s\#Q)^rds\right)^{1/r}\\
&= \left(\int_{\mathcal{S}}W_r(c_s\#Q,c_s\#P)^rds\right)^{1/r}\\
&= FSW_r(Q,P)
\end{align*}
The second line holds by the symmetry property of the ordinary WD. Next, the triangle inequality:
\begin{align*}
FSW_r(P,U) &= \left(\int_{\mathcal{S}}W_r(c_s\#P,c_s\#U)^rds\right)^{1/r}\\
&\leq\left(\int_{\mathcal{S}}\left[W_r(c_s\#P,c_s\#Q)+W_r(c_s\#Q,c_s\#U)\right]^rds\right)^{1/r}\\
&\leq\left(\int_{\mathcal{S}}W_r(c_s\#P,c_s\#Q)^rds\right)^{1/r}+\left(\int_{\mathcal{S}}W_r(c_s\#Q,c_s\#U)^rds\right)^{1/r} \\
&= FSW_r(P,Q) + FSW_r(Q,U)
\end{align*}
The second line holds by the triangle inequality property of the ordinary WD and the third line holds by the Minkowski inequality. Lastly, we check the $r$-convexity property. Let $\lambda \in [0,1]$. If we take $P$ as the reference distribution, this property tells us what happens to the functional sliced WD as we interpolate between $Q$ and $U$:
\begin{align*}
FSW_r(P,\lambda Q+(1-\lambda)U) &= \left(\int_{\mathcal{S}}W_r(c_s\#P,c_s\#[\lambda Q+(1-\lambda)U])^rds\right)^{1/r}\\
&= \left(\int_{\mathcal{S}}W_r(c_s\#P,\lambda c_s\#Q+(1-\lambda)c_s\#U)^rds\right)^{1/r}\\
&\leq \left(\int_{\mathcal{S}}\lambda W_r(c_s\#P,c_s\#Q)^r+(1-\lambda)W_r(c_s\#P,c_s\#U)^rds\right)^{1/r}\\
&\leq\left(\lambda\int_{\mathcal{S}}W_r(c_s\#P,\lambda c_s\#Q)^rds\right)^{1/r}
+\left((1-\lambda)\int_{\mathcal{S}}W_r(c_s\#P,c_s\#U)^rds\right)^{1/r}\\
&= \lambda^{1/r} FSW_r(P,Q) + (1-\lambda)^{1/r} FSW_r(P,U)
\end{align*}
The second line follows from properties of linear operators (in this case $c_s$). The third line follows by the $r$-convexity of the ordinary WD, and the fourth line follows from the Minkowski inequality. We have shown all three pseuduometric properties and the $r-$convexity property, so our proof is complete.
\end{proof}

\newpage
\section{SCWD Implementation}
\label{app:implementation}


The following algorithm describes the process by which we calculated the SCWD values shown in Section $\ref{sec:experiments}$.

\begin{algorithm}
\caption{Spherical Convolutional Wasserstein Distance Approximation}
\label{alg:scwd}
\begin{algorithmic}
\HEADER{Data}
\STATE Reference dataset $X_0(t), t\in \mathcal{T}_0$: sample of spatial fields
\STATE Model outputs $X_1(t), t\in \mathcal{T}_1,...,X_n(t), t\in \mathcal{T}_n$: $n$ samples of spatial fields to be compared to $X_0$
\ENDHEADER

\HEADER{Parameters and Approximation Grids}
\STATE Wasserstein order parameter $r$: 2
\STATE Range parameter $l$: 1,000 km kernel radius 
\STATE Approximation quantiles $Q$: 200 evenly spaced quantiles ranging from 0 to 1 with a step size of 0.005
\STATE Grid $G_1$: $60\times 120$ regular latitude-longitude grid of center points for strided convolution
\STATE Grid $G_2$: $361\times 720$ regular latitude-longitude grid to provide discrete approximation of spherical domain
\ENDHEADER

\HEADER{Step 1. Precompute Slicing Weights}
\FOR{each location $s \in G_1$}
\STATE Calculate vector of chordal distances from $s$ to all locations in $G_2$
\STATE Calculate Wendland function (\ref{eq:wendland}) with range $l$ for all locations in $G_2$ using distance vector
\STATE Apply area weighting to the Wendland kernel values using the area of each grid cell in $G_2$
\STATE Normalize area-weighted kernel and store the results as a sparse vector $W(s)$
\ENDFOR
\ENDHEADER

\HEADER{Step 2. Compute Sliced Quantiles}
\FOR{$i \in 0,1,...,n$}
\STATE Re-grid $X_i$ to $G_2$ without smoothing (one nearest neighbor upsampling)
\FOR{each location $s\in G_1$}
\STATE Slice $X_i$ into one dimension using the dot product $X_i^*(s,t) = \langle W(s), X_i(t)\rangle$
\STATE Calculate the sliced quantile function of $X_i^*(s,t)$, denoted as $F_i^{-1}(s,q), q\in Q$
\ENDFOR
\ENDFOR
\ENDHEADER
\HEADER{Step 3. Calculate Approximate SCWD}
\FOR{$i \in 1,...,n$}
\FOR{each location $s\in G_1$}
\STATE Calculate the local WD between $X_0$ and $X_i$ centered around $s$ as $d_i(s)^r = \sum_{q\in Q}|F_0^{-1}(q)-F_i^{-1}(q)|^r$
\ENDFOR
\STATE Calculate the approximate SCWD between $X_0$ and $X_i$ as $SCWD(X_0,X_i) \approx \left(\sum_{s\in G_1}d_i(s)^r\right)^{1/r}$
\ENDFOR
\ENDHEADER
\end{algorithmic}
\end{algorithm}

The climate model outputs and reanalysis datasets (ERA, NCEP) used in our analysis have no missing data. 
However, the GPCP observational dataset used as the reference for total precipitation did have missing data at some sites for a few days in the historical period.
To handle missing data in the GPCP dataset, we modified the slicing process in Step 2. 
If locations corresponding to greater than 50\% of the convolution weight were missing when calculating a slice value, an \verb|NA| value was recorded. 
The local WD calculations in Step 3 were computed ignoring these \verb|NA| values. In total, 17,943 slices were missing sufficient data when using the 50\% threshold. 
However, a total of 24,105,600 slices were considered for the GPCP dataset (3,348 days in the analysis period with 7,200 slices per day in the strided convolution), so the missing slices constituted under 0.1\% of the final sliced GPCP data used in the analysis.

\newpage
\section{Table of Data Details and Access Links}
\label{sec:data}

\begin{table}[H]
\begin{minipage}[t]{0.5\textwidth}\vspace{0pt}%
\centering
\small
\begin{tabular}{lrrll}
  \hline
Obs./Reanalysis Data & Longs & Lats & TAS & PR \\ 
  \hline
NCEP Reanalysis & 144 &  73 & Yes & Yes \\ 
  ERA5 Reanalysis & 1440 & 721 & Yes & Yes \\ 
  GPCP Observations & 360 & 180 & No & Yes \\ 
   \hline
\end{tabular}

\vspace{6mm}

\begin{tabular}{lrrll}
  \hline
CMIP5 Models & Longs & Lats & TAS & PR \\ 
  \hline
  ACCESS1-0 & 192 & 145 & Yes & Yes \\ 
  ACCESS1-3 & 192 & 145 & Yes & Yes \\ 
  CanCM4 & 128 &  64 & No & Yes \\ 
  CMCC-CESM &  96 &  48 & Yes & Yes \\ 
  CMCC-CM & 480 & 240 & Yes & Yes \\ 
  CMCC-CMS & 192 &  96 & Yes & Yes \\ 
  CNRM-CM5 & 256 & 128 & Yes & Yes \\ 
  CSIRO-Mk3-6-0 & 192 &  96 & Yes & Yes \\ 
  CanESM2 & 128 &  64 & Yes & Yes \\ 
  EC-EARTH & 320 & 160 & Yes & Yes \\ 
  FGOALS-g2 & 128 &  60 & Yes & Yes \\ 
  FGOALS-s2 & 128 & 108 & Yes & Yes \\ 
  GFDL-CM3 & 144 &  90 & Yes & Yes \\ 
  GFDL-ESM2G & 144 &  90 & Yes & Yes \\ 
  GFDL-ESM2M & 144 &  90 & Yes & Yes \\ 
  HadCM3 &  96 &  73 & Yes & Yes \\ 
  HadGEM2-AO & 192 & 145 & Yes & Yes \\ 
  HadGEM2-CC & 192 & 145 & Yes & Yes \\ 
  HadGEM2-ES & 192 & 145 & Yes & Yes \\ 
  INMCM4 & 180 & 120 & Yes & Yes \\ 
  IPSL-CM5A-LR &  96 &  96 & Yes & Yes \\ 
  IPSL-CM5A-MR & 144 & 143 & Yes & Yes \\ 
  IPSL-CM5B-LR &  96 &  96 & Yes & Yes \\ 
  MIROC-ESM & 128 &  64 & Yes & Yes \\ 
  MIROC-ESM-CHEM & 128 &  64 & Yes & Yes \\ 
  MIROC4h & 640 & 320 & Yes & Yes \\ 
  MIROC5 & 256 & 128 & Yes & Yes \\ 
  MPI-ESM-LR & 192 &  96 & Yes & Yes \\ 
  MPI-ESM-MR & 192 &  96 & Yes & Yes \\ 
  MPI-ESM-P & 192 &  96 & Yes & Yes \\ 
  MRI-CGCM3 & 320 & 160 & Yes & Yes \\ 
  MRI-ESM1 & 320 & 160 & Yes & Yes \\ 
  NorESM1-M & 144 &  96 & Yes & Yes \\ 
   \hline
\end{tabular}
\end{minipage}%
\begin{minipage}[t]{0.5\textwidth}\vspace{0pt}%
\centering
\small
\begin{tabular}{lrrll}
  \hline
CMIP6 Models & Longs & Lats & TAS & PR \\ 
  \hline
ACCESS-CM2 & 192 & 144 & Yes & Yes \\ 
  ACCESS-ESM1-5 & 192 & 145 & Yes & Yes \\ 
  AWI-CM-1-1-MR & 384 & 192 & Yes & Yes \\ 
  AWI-ESM-1-1-LR & 192 &  96 & Yes & Yes \\ 
  BCC-ESM1 & 128 &  64 & Yes & Yes \\ 
  CESM2 & 288 & 192 & Yes & Yes \\ 
  CESM2-FV2 & 144 &  96 & Yes & Yes \\ 
  CESM2-WACCM & 288 & 192 & Yes & Yes \\ 
  CESM2-WACCM-FV2 & 144 &  96 & Yes & Yes \\ 
  CMCC-CM2-HR4 & 288 & 192 & Yes & Yes \\ 
  CMCC-CM2-SR5 & 288 & 192 & Yes & Yes \\ 
  CMCC-ESM2 & 288 & 192 & Yes & Yes \\ 
  CanESM5 & 128 &  64 & Yes & Yes \\ 
  E3SM-1-0 & 360 & 180 & Yes & Yes \\ 
  E3SM-2-0 & 360 & 180 & Yes & Yes \\ 
  E3SM-2-0-NARRM & 360 & 180 & Yes & Yes \\ 
  EC-Earth3 & 512 & 256 & Yes & Yes \\ 
  EC-Earth3-AerChem & 512 & 256 & Yes & Yes \\ 
  EC-Earth3-CC & 512 & 256 & Yes & Yes \\ 
  EC-Earth3-Veg & 512 & 256 & Yes & Yes \\ 
  EC-Earth3-Veg-LR & 320 & 160 & Yes & Yes \\ 
  FGOALS-f3-L & 288 & 180 & Yes & Yes \\ 
  FGOALS-g3 & 180 &  80 & Yes & Yes \\ 
  GFDL-CM4 & 288 & 180 & Yes & Yes \\ 
  GFDL-ESM4 & 288 & 180 & Yes & Yes \\ 
  GISS-E2-2-G & 144 &  90 & Yes & Yes \\ 
  ICON-ESM-LR* & N/A & N/A & Yes & Yes \\ 
  IITM-ESM & 192 &  94 & Yes & Yes \\ 
  INM-CM4-8 & 180 & 120 & Yes & Yes \\ 
  INM-CM5-0 & 180 & 120 & Yes & Yes \\ 
  IPSL-CM5A2-INCA &  96 &  96 & Yes & Yes \\ 
  IPSL-CM6A-LR & 144 & 143 & Yes & Yes \\ 
  IPSL-CM6A-LR-INCA & 144 & 143 & No & Yes \\ 
  KACE-1-0-G & 192 & 144 & Yes & Yes \\ 
  KIOST-ESM & 192 &  96 & Yes & Yes \\ 
  MIROC6 & 256 & 128 & Yes & Yes \\ 
  MPI-ESM-1-2-HAM & 192 &  96 & Yes & Yes \\ 
  MPI-ESM1-2-HR & 384 & 192 & Yes & Yes \\ 
  MPI-ESM1-2-LR & 192 &  96 & Yes & Yes \\ 
  MRI-ESM2-0 & 320 & 160 & Yes & Yes \\ 
  NESM3 & 192 &  96 & Yes & Yes \\ 
  NorCPM1 & 144 &  96 & Yes & Yes \\ 
  NorESM2-LM & 144 &  96 & Yes & Yes \\ 
  NorESM2-MM & 288 & 192 & Yes & Yes \\ 
  SAM0-UNICON & 288 & 192 & Yes & Yes \\ 
  TaiESM1 & 288 & 192 & Yes & Yes \\ 
   \hline
\end{tabular}
\end{minipage}%
\caption{\label{tab:meta} Additional details for each observed/reanalysis data product and CMIP model output. The columns give the model name, longitude and latitude resolution, and availability of (2m) surface temperature (TAS) and total precipitation (PR) for each dataset. All datasets were obtained on a rectangular grid except ICON-ESM-LR, which was obtained on an icosahedral grid with 10,242 total cells.}
\end{table}

CMIP5 and CMIP6 outputs: \url{https://esgf-node.llnl.gov/projects/esgf-llnl/}

ERA5 hourly data on single levels from 1940 to present: \url{https://cds.climate.copernicus.eu/cdsapp#!/dataset/reanalysis-era5-single-levels?tab=overview}

NCEP/DOE Reanalysis II: \url{https://psl.noaa.gov/data/gridded/data.ncep.reanalysis2.html}. 

GPCP 1 Degree Daily Precipitation Estimate: \url{https://www.ncei.noaa.gov/products/climate-data-records/precipitation-gpcp-daily}. 

\newpage
\section{Full SCWD Rankings and Metric Comparison}
\label{app:tables}

\begin{table}[H]
\centering
\includegraphics[width=7in]{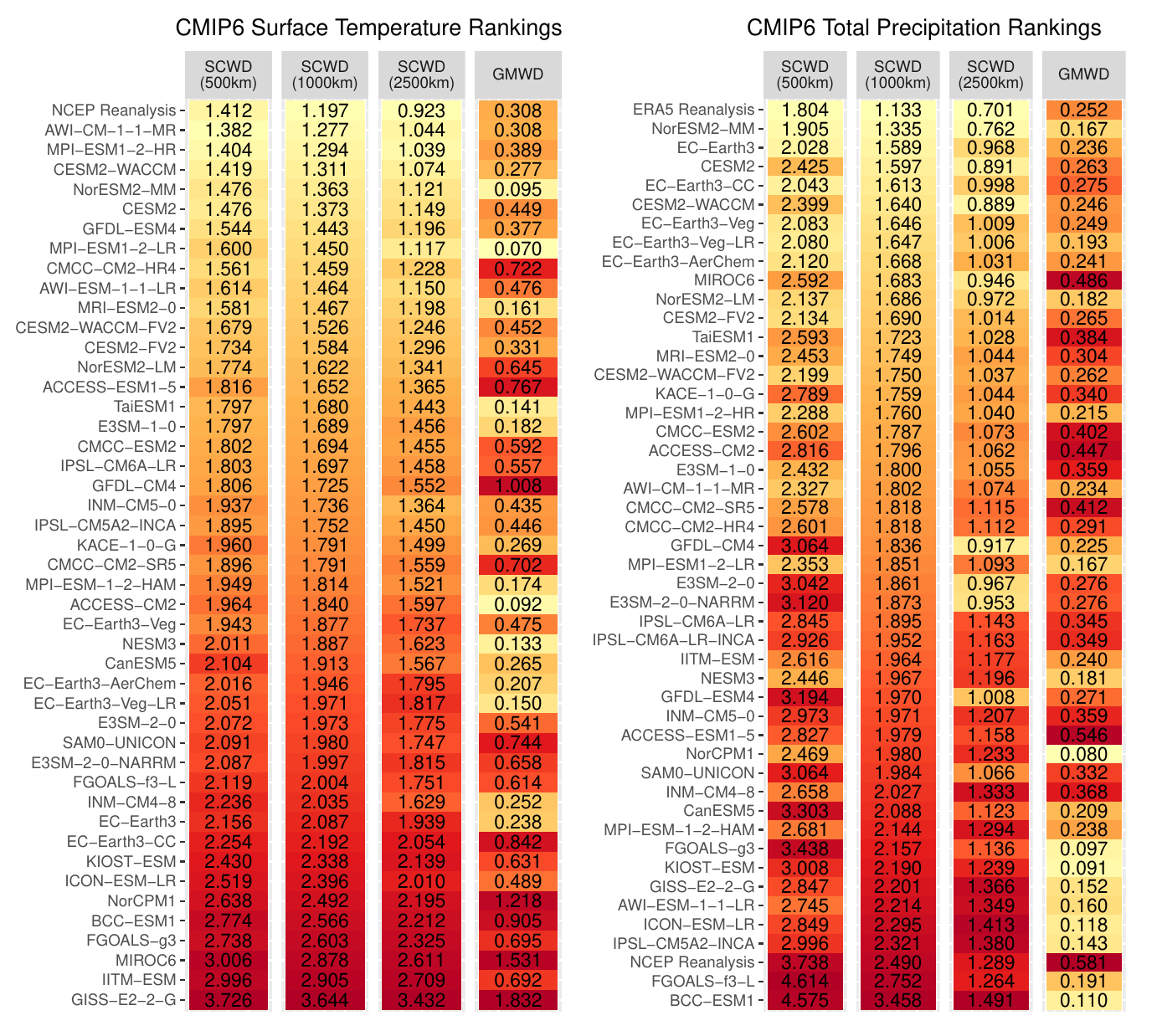}
\caption{CMIP6 model rankings for both (2m) surface temperature and total precipitation based on similarity to the ERA5 Reanalysis (surface temperature) and GPCP observations (total precipitation). Distances are calculated using our proposed spherical convolutional WD (SCWD) as well as the global mean-based WD (GMWD) from \cite{vissio2020evaluating}. For the SCWD calculations, three different range parameters are chosen for the Wendland kernel: 500km, 1000km (our proposed choice), and 2500km. Color fill is unique to each column in the tables, and is calculated using ranks.}
\label{tab:cmip6}
\end{table}

\begin{table}[h]
\centering
\includegraphics[width=7in]{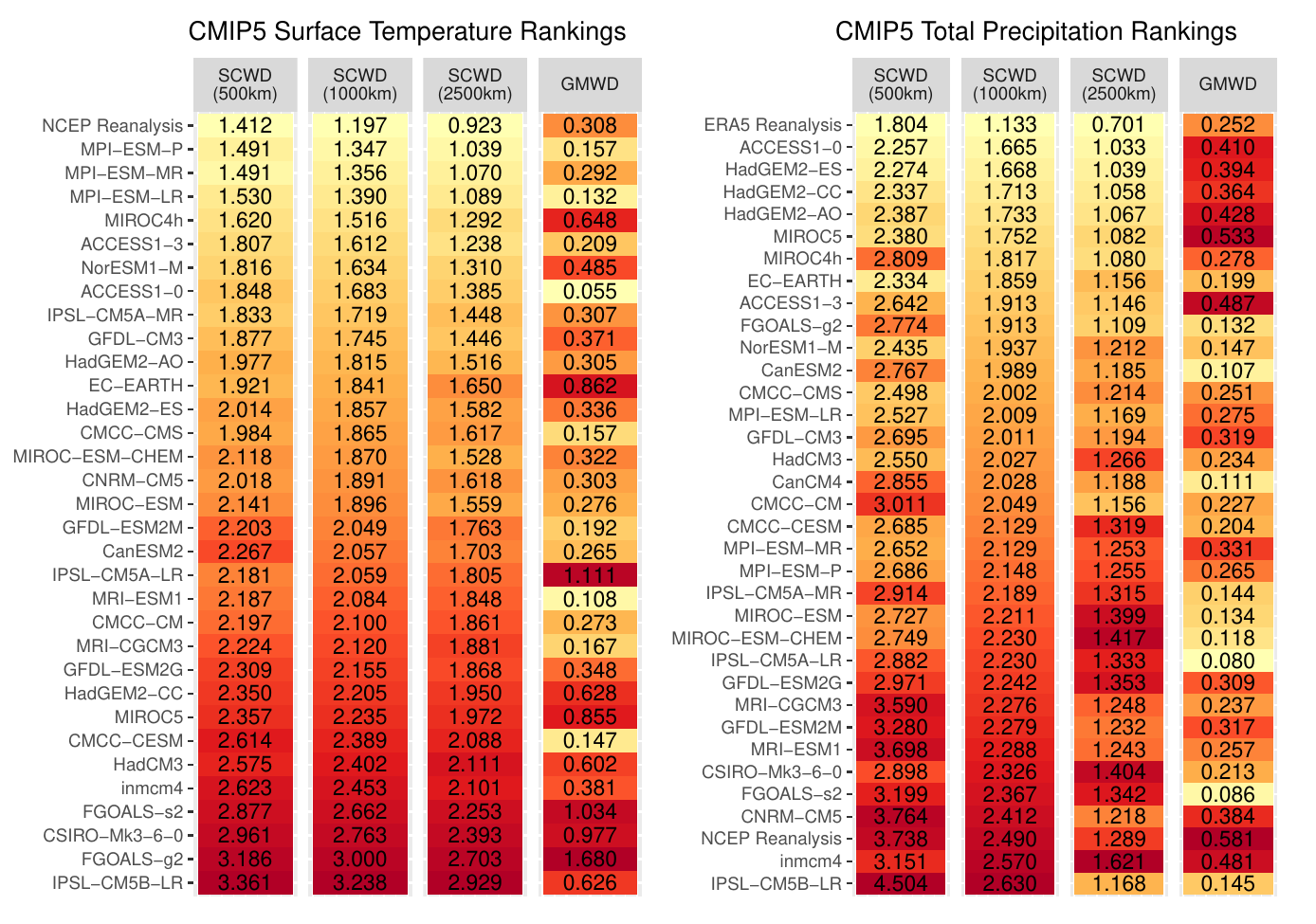}
\caption{CMIP5 model rankings for both (2m) surface temperature and total precipitation based on similarity to the ERA5 Reanalysis (surface temperature) and GPCP observations (total precipitation). Distances are calculated using our proposed spherical convolutional WD (SCWD) as well as the global mean-based WD (GMWD) from \cite{vissio2020evaluating}. For the SCWD calculations, three different range parameters are chosen for the Wendland kernel: 500km, 1000km (our proposed choice), and 2500km. Color fill is unique to each column in the tables, and is calculated using ranks.}
\label{tab:cmip5}
\end{table}




\end{document}